\documentclass[12pt,osajnl2,twocolumn,showpacs,superscriptaddress]{revtex4}  %% REVTeX 4.0
\usepackage[draft]{hyperref} %% optional
\begin{document}

%\twocolumn[
\title{Circular dichroism in planar nonchiral plasmonic metamaterials}

\author{Vassilios Yannopapas}
\address{Department of Materials Science, School of Natural Sciences, \\
         University of Patras, GR-26504 Patras, Greece}

\begin{abstract}
It is shown theoretically that a nonchiral, two-dimensional array
of metallic spheres exhibits optical activity as manifested in
calculations of circular dichroism. The metallic spheres occupy
the sites of a rectangular lattice and for off-normal incidence
they show a strong circular-dichroism effect around the surface
plasmon frequencies. The optical activity is a result of the
rectangular symmetry of the lattice which gives rise to different
polarizations modes of the crystal along the two orthogonal
primitive lattice vectors. These two polarization modes result in
a net polar vector, which forms a chiral triad with the wavevector
and the vector normal to the plane of spheres. The formation of
this chiral triad is responsible for the observed circular
dichroism, although the structure itself is intrinsically
nonchiral.
\end{abstract}

\maketitle

Author's email address is: vyannop@upatras.gr

% \section*{ACKNOWLEDGEMENTS}
%This work was supported by the `Karatheodory' research fund of
%University of Patras.

Metamaterials is new class of artificial materials possessing
properties and capabilities which are not met in their constituent
(naturally occurring) materials. The majority of the recent work
in this field focuses on the achievement of subwavelength
structures with negative refractive index (NRI). This can be
achieved when the effective permittivity and permeability of the
metamaterial become simultaneously negative in a common spectral
region \cite{pendry,smith2}. In an alternative route, a NRI can
occur in a chiral material consisting of resonant elements
\cite{pendry_chiral}. There have been several theoretical
proposals \cite{lakhtakia_1,tretyakov_1,lakhtakia_2,
lakhtakia_3,monzon,tretyakov_2,shen,agra,yanno_chiral,baev,kwon}
of artificial chiral metamaterials exhibiting NRI but only
recently there has been experimental confirmation of
chirality-assisted NRI \cite{rogacheva,dong}. In this work, the
optical activity and the resulting NRI stem either from the
helical symmetry of the elements repeated periodically in the
metamaterial or from the helical symmetry of the underlying
three-dimensional (3D) lattice. Recently, strong optical activity
has been reported in a two-dimensional (2D) nonchiral metamaterial
\cite{fedotov,plum}. The latter was realized as a perforated thin
metallic sheet consisting of a square array of pairs of arcs of
different length. This structure exhibited strong circular
dichroism for off normal incidence of circular plane waves and in
the case where the plane of incidence did not contain the polar
vector of the unit cell. The polar vector occurred from the
asymmetry of the unit cell (arcs of unequal length); this vector
along with the wavevector and the normal to the metamaterial sheet
form a chiral triad resulting in the observed optical activity.
The above effect has been termed as extrinsic chirality
\cite{fedotov} due to the nonchiral nature of the studied
structures.

In the present Letter we employ a relatively simpler nonchiral 2D
metamaterial exhibiting strong circular dichroism. The
metamaterial consists of metallic (plasma) spheres occupying the
sites of rectangular lattice. A rectangular lattice is, by
definition, a 2D anisotropic system which possesses two principal
polarization directions. Under plane wave illumination and around
the dipolar surface plasmon resonance, the above structure
exhibits two distinct absorption peaks corresponding to the two
different polarization modes resulting from the anisotropy of the
lattice. For oblique incidence and circular polarization, the
absorbance of the metamaterial depends on the state of
polarization (left or right polarization) as long as the two
principal axes of the metamaterial are not contained in the plane
of incidence. The metamaterial under study operates in the optical
regime where the optical activity is manifested within the
surface-plasmon band of the metallic spheres. As such, it
constitutes a much more realistic proposal of a planar, optically
active metamaterial in the visible regime than the miniaturization
of the corresponding ones of Refs.~\onlinecite{plum} and
\onlinecite{fedotov}. Furthermore, the inclusion of plasmonic
materials (metals) in artificial chiral structures greatly
enhances gyrotropic effects when compared with naturally occurring
chiral materials \cite{schwan1,schwan2,decker}.

\begin{figure}[htbp]
\centering
\includegraphics[width=8.3cm]{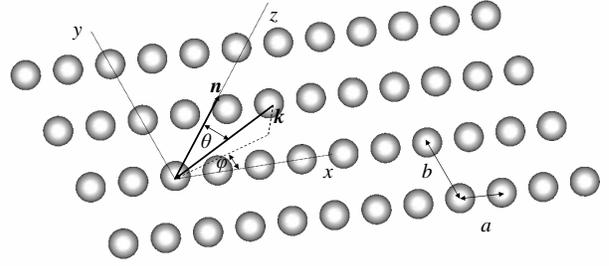}
\caption{Rectangular array of plasma spheres where the sides of
the unit cell are denoted by $a$ and $b$, respectively. $\theta$
is the angle between the incident wavevector ${\bf k}$ and the
vector ${\bf n}$ which is normal to the plane of spheres
($xy$-plane). $\phi$ is the angle between the parallel (to the
$xy$-plane) component of the wavevector and the $x$-axis. }
\label{fig1}
\end{figure}

Fig.~\ref{fig1} shows the plasmonic metamaterial under study: a
lattice with a rectangular unit cell with sides $a$ and $b$,
occupied by metallic spheres described by a Drude-type dielectric
function, i.e., $\epsilon(\omega)=1-\omega_{p}^{2}/[\omega (\omega
+ {\rm i} \gamma)]$, where $\omega_{p}$ is the plasma frequency of
the metal and $\gamma$ is the loss factor which is taken to be
$\gamma =0.01 \omega_{p}$, a typical value for metallic
nanoparticles. The radius of the plasma spheres is taken to be
$S=0.3 c/ \omega_{p}$. The optical properties of the above
structure have been studied by means of the
layer-multiple-scattering (LMS) method which is based on an ab
initio multiple-scattering theory, using a well documented
computer code \cite{skm92,comphy}. The LMS method is ideally
suited for the calculation of the transmission, reflection and
absorption coefficients of an electromagnetic (EM) wave incident
on a composite slab consisting of a number of layers which can be
either planes of non-overlapping particles with the same 2D
periodicity or homogeneous plates. For each plane of particles,
the method calculates the full multipole expansion of the total
multiply scattered wave field and deduces the corresponding
transmission and reflection matrices in the plane-wave basis. The
transmission and reflection matrices of the composite slab are
evaluated from those of the constituent layers.

\begin{figure}[htbp]
\centering
\includegraphics[width=8.3cm]{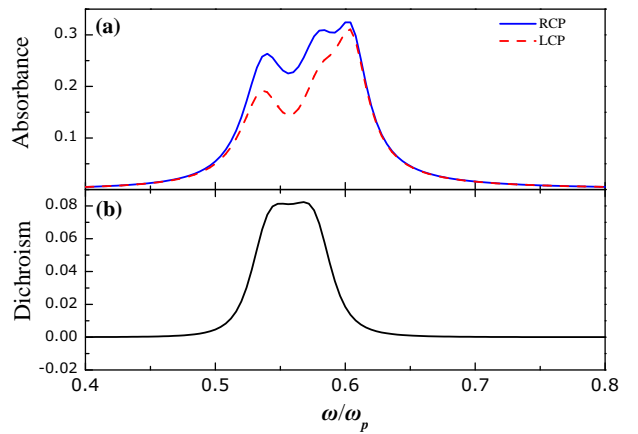}
\caption{(Color online) (a) Absorbance spectra for right- (solid
line) and left-circularly (broken line) polarized light incident
at angles $\theta=\phi=45^{0}$ (see Fig.~\ref{fig1} for their
definition) on a rectangular ($a=b/2=c/ \omega_{p}$) array of
plasma spheres ($S=0.3 c/ \omega_{p}$). (b) The corresponding
circular dichroism spectrum of the above system of plasma
spheres.} \label{fig2}
\end{figure}

In what follows we study the absorption spectra for right- and
left-circularly polarized plane waves, denoted by $A_{R}$ and
$A_{L}$. The corresponding dichroism is given by the difference of
the above absorption spectra, i.e., $\Delta=A_{R}-A_{L}$. We have
considered the general case of oblique incidence where the
incident wavevector is defined by the angles $\theta$ and $\phi$
shown in Fig.~\ref{fig1}. Fig.~\ref{fig2}a shows the absorbance
for left- and right-circularly polarized (RCP-solid curve and
LCP-broken curve, respectively) incident light with
$\theta=\phi=45^{0}$.  Both RCP and LCP curves exhibit two main
absorption bands corresponding to the two different dipolar EM
modes stemming from the coupling of neighboring spheres along the
two principal axes of the lattice. We note that under circularly
polarized plane-wave illumination, both modes are excited. In the
case of normally incident wave polarized along one of the
principal axis, only one of the modes is excited and a single,
dipolar peak is observed \cite{ys_2004}. The high-frequency
absorption band possesses an inner structure which stems from the
excitation of higher-multipole surface plasmons \cite{ys_2004}.
The main finding of Fig.~\ref{fig2}a is the fact that the light
absorption strongly depends on the state of circular polarization
of the incident wave; this is also manifested in the dichroism
curve of Fig.~\ref{fig2}b. We note that the dichroism vanishes for
normal incidence as well as for oblique incidence where
$\phi=0^{0}$ or $90^{0}$ in which case the parallel (to the
surface) component of the wavevector coincides with one of the two
principal directions of the rectangular lattice. This is due to
the fact that, in order to achieve circular dichroism in our case,
the direction of the incident light must constitute a ``screw
direction'' of the unit cell, i.e. the wavevector, the vector
normal to the surface and one of the two primitive lattice vectors
must form a chiral triad \cite{bunn,fedotov,plum}.

\begin{figure}[htbp]
\centering
\includegraphics[width=8.3cm]{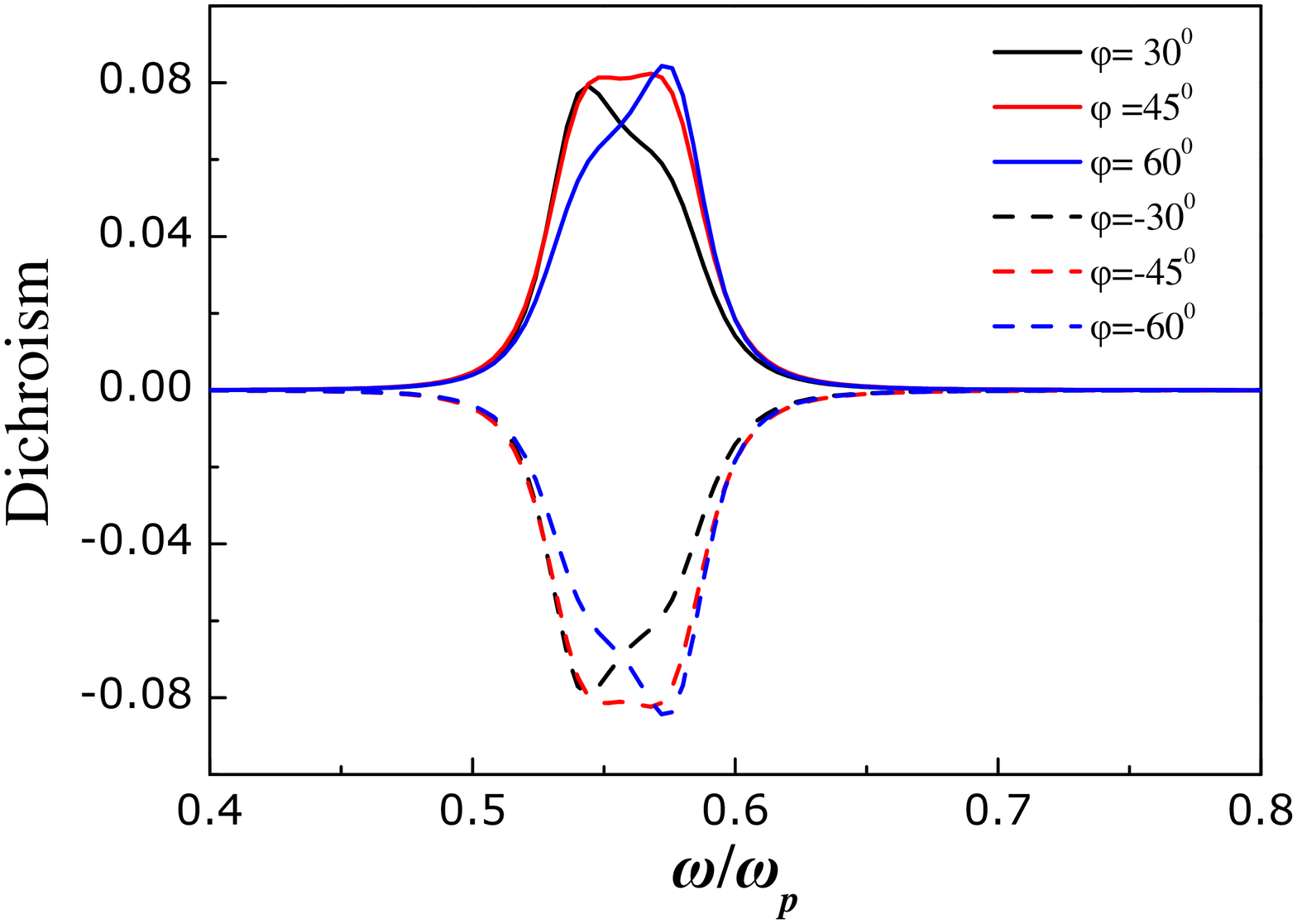}
\caption{(Color online) Circular dichroism spectra for
$\theta=45^{0}$ and for various values of the angle $\phi$ (shown
in the inset) for the array of plasma spheres of Fig.~\ref{fig2}.}
\label{fig3}
\end{figure}

\begin{figure}[htbp]
\centering
\includegraphics[width=8.3cm]{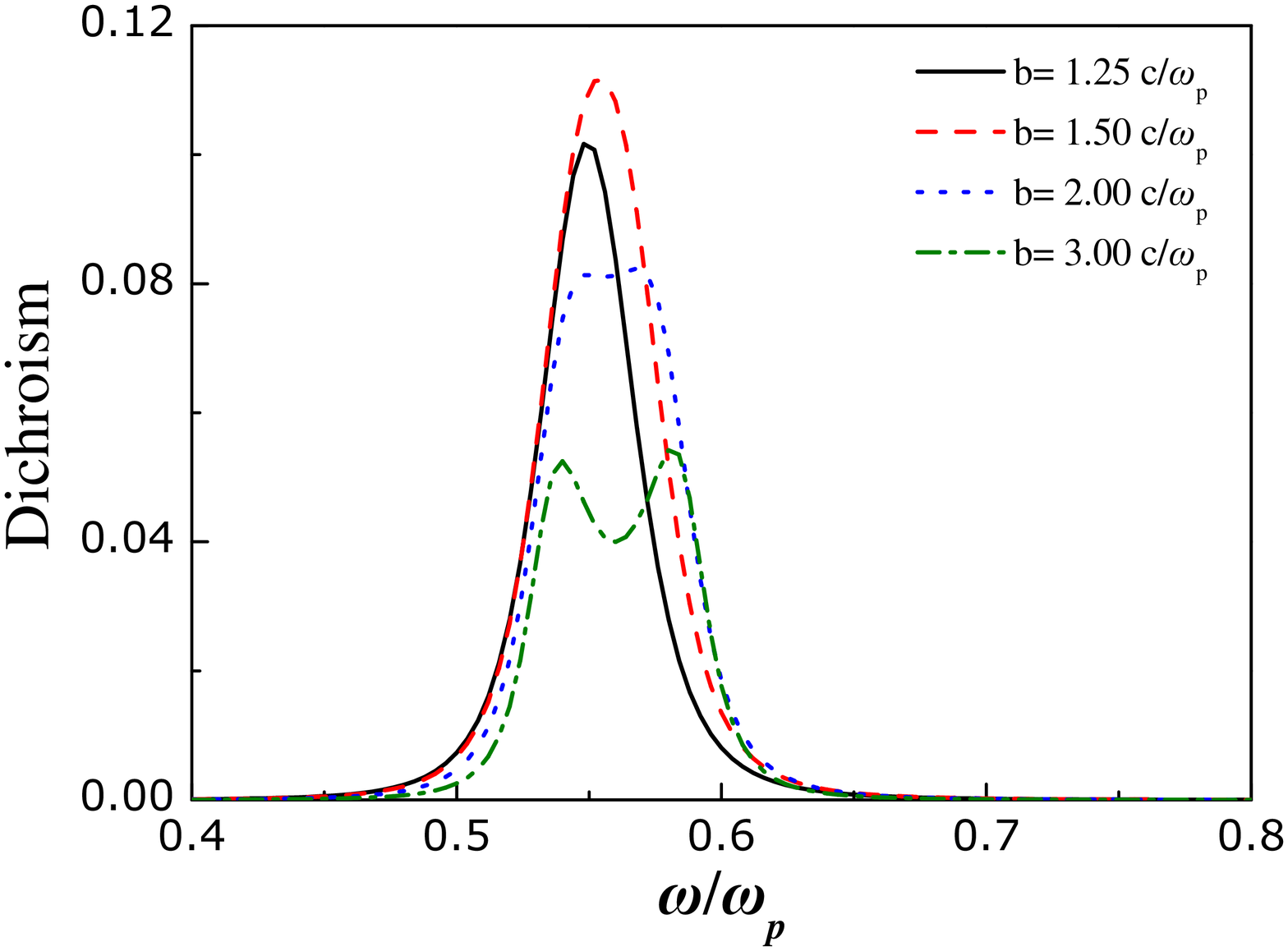}
\caption{(Color online) Circular dichroism spectra for
$\theta=\phi=45^{0}$ and different rectangular arrays of plasma
spheres ($S=0.3 c/ \omega_{p}$) with $a=c/ \omega_{p}$ and $b$ as
shown in the inset.} \label{fig4}
\end{figure}

In Fig.~\ref{fig3} we show circular dichroism spectra for a fixed
angle $\theta=45^{0}$ and for different values of $\phi$. As
expected, the dichroism spectra have opposite signs for opposite
values of $\phi$ which correspond to two enantiomeric
arrangements. In general, the spectral bands of circular dichroism
for the different values of $\phi$ have more or less the same
width and height. Fig.~\ref{fig4} depicts optical activity spectra
for fixed $\theta=\phi=45^{0}$ and for different values of the
side $b$ of the rectangular unit cell. Since the optical activity
is lost for a square array of spheres, as $b/a$ increases from
unity, we expect the circular dichroism to increase. This is
evident by comparing the curves for $b=1.25 a$ and $b=1.5 a$.
However, for larger values of $b$, the trend is reversed and the
optical activity is suppressed. This is due to the fact that as
$b$ increases, the lattice becomes very dilute, the coupling of
neighboring spheres in the $y$-direction diminishes and so does
the light absorption (for both modes of circular polarization).

In conclusion, it has been shown that a nonchiral planar lattice
of metallic spheres exhibits strong optical activity around the
surface plasmon frequencies, under suitable conditions of light
incidence. The structure can be fabricated by contemporary
fabrication techniques \cite{taleb} and used as a compact
optically-active component in nanostructured devices.

\end{document}